# Competition between covalent bonding and charge transfer at complex-oxide interfaces

Juan Salafranca,[1,2,*] Julián Rincón,[2,3] Javier Tornos,[1] Carlos León,[1] Jacobo Santamaria,[1] Elbio Dagotto,[2,4] Stephen J. Pennycook,[5,2] and Maria Varela.[2,1]

[1]Grupo de Física de Materiales Complejos, Universidad Complutense, 28040 Madrid, Spain

[2]Materials Science and Technology Division, Oak Ridge National Laboratory, Oak Ridge, TN 37831, USA

[3]Center for Nanophase Materials Sciences, Oak Ridge National Laboratory, Oak Ridge, TN 37831, USA

[4]Department of Physics and Astronomy, The University of Tennessee, Knoxville, TN 37996, USA

[5]Department of Materials Science and Engineering, The University of Tennessee, Knoxville, TN 37996, USA

*Correspondence to: jsalafra@ucm.es

Here we study the electronic properties of cuprate/manganite interfaces. By means of atomic resolution electron microscopy and spectroscopy, we produce a subnanometer scale map of the transition metal oxidation state profile across the interface between the high Tc superconductor $YBa_2Cu_3O_{7-\delta}$ and the colossal magnetoresistance compound $(La,Ca)MnO_3$. A net transfer of electrons from manganite to cuprate with a peculiar non-monotonic charge profile is observed. Model calculations rationalize the profile in terms of the competition between standard charge transfer tendencies (due to band mismatch), strong chemical bonding effects across the interface, and Cu substitution into the Mn lattice, with different characteristic length scales.

A detailed understanding of the charge transfer that occurs across semiconductor interfaces has led to the development of two-dimensional electron gases[1], as well as the integer and fractional quantum Hall effect.[2–4] Interfaces between transition-metal oxides (TMO's) have the potential for even richer physics, due to the presence of several competing interactions with similar characteristic energies. The competition between electrostatic effects – similar to those at work in semiconductor heterostructures – and orbital physics characteristic of TMO's can give rise to exotic electronic reconstructions and novel physical behaviors. In heterostructures of $LaAlO_3/SrTiO_3$, the observation of a metal-insulator transition at the interface of these non-magnetic (bulk) insulators[5] (along with superconductivity[6] and magnetism[7]) sparked considerable interest. However, oxide interfaces also bring along many challenges. Ionic defects such as oxygen vacancies might play an important role in determining the electronic structure.[8-13] Understanding and controlling these material-physics issues – and the effect they have on the properties – is essential to fully explore the new functionalities that these fascinating compounds might bring along.[14]

Ferromagnetic/superconducting interfaces of $La_{2/3}Ca_{1/3}MnO_3/YBa_2Cu_3O_{7-\delta}$ (LCMO/YBCO) have attracted much attention. This system is a paradigmatic example of competition between strongly correlated systems with different ground states. It has been proposed, based on the difference between chemical potentials, that electronic charge would be transferred from the manganite to the cuprate.[15,16] This mechanism, however, does not consider the details of the interface. The interfacial electronic structure depends on other details, such as the atomic termination[17] for each material. At the LCMO/YBCO interface both a change in the orbital occupation and a net magnetic moment are induced in the cuprate.[18,19] Model calculations[20] were able to explain different experimental results regarding the competition between ferromagnetism and superconductivity[21]. However, the effect of charge transfer was not studied. Very recently, cross-sectional scanning tunneling microscopy measurements have suggested[22] that charge transfer takes place with a characteristic length scale of ~1 nm. However, the interpretation of these measurements is unclear. Further work aimed at studying the electronic structure – including charge distributions – and the importance of interface and bulk effects is necessary to gain full understanding of properties of these interfaces.

In this letter, we present a combined experimental and theoretical study of the (100) LCMO/YBCO interface. The unique capabilities of scanning transmission electron microscopy (STEM), in combination with electron energy-loss spectroscopy (EELS), allow us to identify the precise chemical terminations, and to establish an oxidation state profile with sub-nanometer resolution. We find an anomalous charge redistribution, with a non-monotonic behavior of the occupancy of d-orbitals in the manganite layers, as a function of distance to the interface. Model calculations indicate that this profile is a result of the competition between standard charge transfer tendencies, strong bonding effects across the interface, and Cu substitution into the Mn lattice. We also study the effect of oxygen vacancies, electron-electron interactions, and the polar discontinuity mechanism, and we find that their effect is not important in reproducing the shape of the charge distribution.

A high magnification Z-contrast image of a $La_{0.7}Ca_{0.3}MnO_3/YBCO/La_{0.7}Ca_{0.3}MnO_3$ trilayer[23] is shown in Fig. 1(a). Interfaces are sharp and coherent, and the structural quality of the samples is high. Occasionally, interface steps one unit cell high are observed.[24] The structure is, however, unusual near the interfaces. Fig. 1(b) shows the 3d metal-to-3d metal distance along the growth direction (z) for the whole image While the lattice parameter in the manganite is constant all the way to the interface, the YBCO intracell distances exhibit a non-linear relaxation with a characteristic length of one or two unit cells. The $CuO_2$ planes in the first and second unit cells by the interface move further apart from each other, while the $CuO_2$ plane to CuO chain distance is somewhat decreased. These non-linear effects are likely related to the relaxation of epitaxial strain due to lattice mismatch.[25] Also, both top and bottom interfaces lack CuO chains (this atomic plane exhibits the darkest contrast).[24] Confirmation of the stacking sequence can be obtained by EEL spectrum images. Figure 1(c) shows atomic resolution maps of the O K, Mn $L_{2,3}$, Ba $M_{4,5}$ and La $M_{4,5}$ absorption edges, respectively. The overlay of Mn (red), La (green), and Ba (blue) maps proofs that at both interfaces a Ba-O plane is facing a Mn-O plane. In the predominant termination, no interfacial CuO chains are observed. Spectroscopic data, including linescans as the one in Fig. 1(c), show that the interfaces are chemically abrupt within the precision of the technique, limited by the unavoidable formation of amorphous layers during specimen preparation (SM). The abrupt interface is consistent with previous x-ray work.[23]

These structural changes have a direct impact on the electronic properties, which can also be analyzed from EELS. The EELS fine structure reflects the details of the unoccupied density of states. In particular, the O K-edge fine structure correlates with the electronic doping in both manganites and cuprates,[21,26] as does the intensity ratio between the $L_2$ and $L_3$ edges of Mn. Figure 2(a) shows the variation in the O K edge across several LCMO (red)/YBCO (blue) bilayers superimposed over a low magnification image of a YBCO/LCMO superlattice. Figure 2(b) shows the actual background subtracted spectra, acquired while moving from the middle of a LCMO layer into the adjacent YBCO layer. Changes both in the intensity of the main peak (≈ 535 eV), the prepeak (≈ 530 eV) and its position (dashed lines) can be observed. The profiles for the prepeak intensity and the position of the edge onset are shown in Fig. 2(c). These quantities are not the same. The changing onset of the absorption edge is produced by the shift in the core-level energies. It is reasonable to assume that the bulk chemical potential of YBCO is around 2 eV lower than the LCMO bulk chemical potential,[15] and a net transfer of electrons from manganite to cuprate takes place until the chemical potentials reach equilibrium, shifting the core levels a similar energy. On the other hand, the prepeak intensity reflects the occupation of specific orbitals, as we present later; it has been found to be proportional to the oxidation state of the transition metal in manganites,[27] and it is also correlated with the hole carrier density in YBCO[26]. Therefore, Fig. 2 reveals both the formal Mn valence within the LCMO layers and the hole doping in the YBCO. Near the interface, the prepeak intensity in YBCO decreases indicating a reduced hole density (i.e., the electron doping increases). The prepeak is also reduced within a nm in the LCMO side of the interface, sign of a reduced Mn oxidation state,[27] also consistent with the sign of the difference in bulk chemical potential.

However, a more refined analysis of the charge profiles in manganite layers of different thicknesses (Fig. 3(a)) reveals surprises. These profiles have been calculated by subtracting the Mn valence measured from the $L_{23}$ intensity ratio[27] from the nominal +3.3 expected according to the chemical doping. A few nanometers away from the interface, LCMO shows a deficit of electrons, as expected to compensate for the extra electrons in YBCO. It is worth noting that these experiments were carried out at room temperature where YBCO is a bad metal, and LCMO is an insulator. Screening in the insulating phase is significantly less efficient, resulting in charge transfer with a much larger characteristic length. The overall profile is compatible with an electron reconstruction driven by the chemical potential mismatch of the two materials.[15] However, the region closest to the interface shows an electron enrichment at both sides of the interface. This unexpected behavior is in principle incompatible with usual semiconductor-like physics, implying the appearance of an additional energy scale competing with charge transfer effects.

In order to explore the origin of the unexpected charge distribution, we turn to modelcalculations. We concentrate on two basic interactions: the kinetic energy of conduction and valence electrons – due to the hybridization of d-like orbitals – and the Coulomb interaction among them and with the ions and core electrons. The effective dielectric constant in YBCO (a metal at room temperature) is chosen much larger than in LCMO (an insulator, details in the supplemental material (SM)). To model the kinetic energy, the two $e_g$ orbitals are important in both manganites[28] and in YBCO near the interface.[18,30] Therefore, we have considered a two-orbital tight binding model with effective hopping and electronic interactions ($t_0 \approx 0.5$eV, the manganite-bulk hopping parameter is taken as the energy unit,[29] see SM). The effective values of the hopping parameter in the z direction across the interface, and in the first manganite layer, $t$ and $t'$, might be strongly affected by interface effects, such as the observed lattice relaxations (Fig 1(b)). They are the most important parameters in this work, because we use them to explore interface effects in the electronic structure and charge distributions.

Let us now consider possible causes for the atypical charge distribution, starting with the polar discontinuity effect that arises at the interface of two materials with different formal polarizations.[9] In this situation, the electric displacement field grows with increasing layer thickness, unless a transfer of charge towards the interface occurs. This effect is implicitly included in our model, where the potential is calculated by assigning to each atomic plane the charge corresponding to the Wannier functions centered in that plane. One way to isolate the effect of polar discontinuity is to get rid of the formal polarization in each unit cell of the different materials. We can do so by substituting all charge in the different unit cells of each material by a point charge with a value that equals the net charge within each unit cell (the exact value determined by the self-consistent calculation). We place these charges in between the $CuO_2$ biplanes, and the $MnO_2$ planes of cuprate and manganite. Then, the Coulomb potential produced does not depend on the particular termination of any material, thus eliminating the effect of polar discontinuity. However, Fig. 3(b) shows that this particular interface termination enriches the LCMO side of the interface with holes (instead of electrons as in the experiments). Therefore, polar discontinuity is insufficient to understand the phenomena discussed here.

Consider now the influence of oxygen vacancies near the interface. The presence of a significant number of oxygen vacancies is unlikely because the samples are grown in a high-oxygen pressure environment.[23] However, oxygen vacancies (difficult to detect) dope the system with electrons. Furthermore, in epitaxial thin films they can help releasing strain. In order to include them in the model, we adjust the formal charge of the first $MnO_2$ plane (see SM) to the charge that corresponds to MnO, while preserving charge neutrality. Polar discontinuity effects are properly included. The resulting charge profile is shown in Fig. 3(c), some general features similar to the experiment are found in the YBCO region; however, there is an important difference in the LCMO region, since the experimental profile has a non-monotonic behavior.

More complex vacancy distributions are possible, but there is a limitation on the effect of vacancy doping. By applying Gauss's law – and assuming translation invariance parallel to the interface –, it is possible to show that whether the energetics favor electrons or holes near the interface further away the electrostatic interactions would make the charge density tend to the bulk value, creating a monotonic profile. This is true regardless of the values of material dependent dielectric constants, which determine the decay lengths of the charge profiles but not the general features. Therefore, electrostatic effects alone are simply unable to reproduce the experimental non-monotonic profile.

We turn our attention to the effect of covalent bonding across the interface (due to the strong overlap between the orbitals at both sides).[30] This effect can be included in the model by increasing the hopping across the interface (t). Additionally, we also consider the changes in the hopping between orbitals in the first two layers of the manganite (t′). An increased hopping across the interface is supported by experiments showing orbital reconstruction,[18] and a strong magnetic coupling between Cu and Mn moments.[19,31] The results in Fig. 3(d), (for $t=10t_0$ $t'=4t_0$) show a non-monotonic charge profile in the manganite layer. Taking also into account possible substitution of Cu into the Mn lattice improves the agreement between experiments and calculations, although small to moderate substitution alone cannot account for the non-monotonic profile by itself (nor can other kinds of chemical disorder, details in SM).

Electron-electron interactions do not alter this picture. A numerically exact treatment of the electron-electron interaction is possible via the density matrix renormalization group (DMRG),[32] although dimensionality is then constrained to one. Figure 3(e) illustrates the results of DMRG for a one-dimensional version of the model described above with the inclusion of an interaction term (details in SM). Both the charge redistribution and Friedel oscillations are strongly suppressed by electron-electron interaction, and for a value of $U=4t_0$, the charge distribution essentially follows the background charge. Thus, the Hubbard U does not play an important role in explaining our experimental results.

The mechanism by which the large hybridization results in an excess of electrons near the interface can be understood in terms of bonding between Cu and Mn orbitals. In the limit of $t \gg t_{Mn}^z, t_{Cu}^z$, a bonding and antibonding orbital will form. The bonding orbital will be occupied making the charge at each of the sites equal to 1⁄2 electron. In our two-orbital model, a large hopping across the interface between two particular orbitals (in this case $3z^2-r^2$ for Cu and Mn) results in a tendency of these orbitals to have a filling close to half an electron per orbital. This explains why holes appear in the $3z^2-r^2$ orbital in YBCO near the interface[15] – normally full in bulk YBCO –, while electrons appear in the $3z^2-r^2$ orbital in LCMO, which normally has 0.33 electrons for the doping considered here.

The results in Fig. 3(d) agree with the experimental profile. However, the theoretical profile filling near the interface is never larger than the bulk filling. There are different possible causes for this discrepancy. In the model, the two active Mn orbitals have a non-zero density of states at the Fermi energy, providing enough freedom to screen the extra charge that finds its way to the bonding orbital. A more elaborate model that is able to reproduce the insulating character of LCMO should therefore lead to a better agreement. Among different types of chemical disorder, calculations indicate that small Cu substitution into the Mn lattice improves the agreement with experiments, if hybridization is also considered (SM). Theoretical and experimental results are overall similar and the mechanism due to hybridization of Cu and Mn orbitals, possibly complemented by a small Cu/Mn substitution in LCMO, provides a rationale for the relative electron enrichment of LCMO near the interface.

In summary, the competition between electronic reconstruction (due to band mismatch of YBCO and LCMO) and the strong bonding across the interface appears responsible for the exotic charge profile observed at YBCO/LCMO interfaces. This competition can be traced down to a combination of electrostatic effects – similar to those at work in semiconductor heterostructures – and orbital physics – characteristic of TMO's. The charge profile and interface physics will depend on the energetics of the $e_g$ levels, and therefore it might be tuned by strain, doping, supeconductivity,[33] and electron-lattice interactions.[14]

Acknowledgments. The authors thank Luis Brey for helpful discussions and Masashi Watanabe for the principal component analysis plug-in for Digital Micrograph. Research at ORNL (SJP, MV, ED, and JR) was supported by the U.S. Department of Energy (DOE), Basic Energy Sciences (BES), Materials Sciences and Engineering Division, and through the Center for Nanophase Materials Sciences (CNMS), which is sponsored by the Scientific User Facilities Division, DOE-BES.  JSal was supported by the ERC starting Investigator Award, grant #239739 STEMOX and Juan de la Cierva program JCI-2011-09428 (MICINN-Spain). Research at UCM (JT, CL, JSan) was supported by the Spanish MICINN/MINECO through grants MAT2011-27470-C02 and Consolider Ingenio 2010 - CSD2009-00013 (Imagine), and by CAM through grant S2009/MAT-1756 (PHAMA). Computations were supported by the National Center for Supercomputing Applications (U.S. Department of Energy, contract no. DE-AC02-05CH11231).

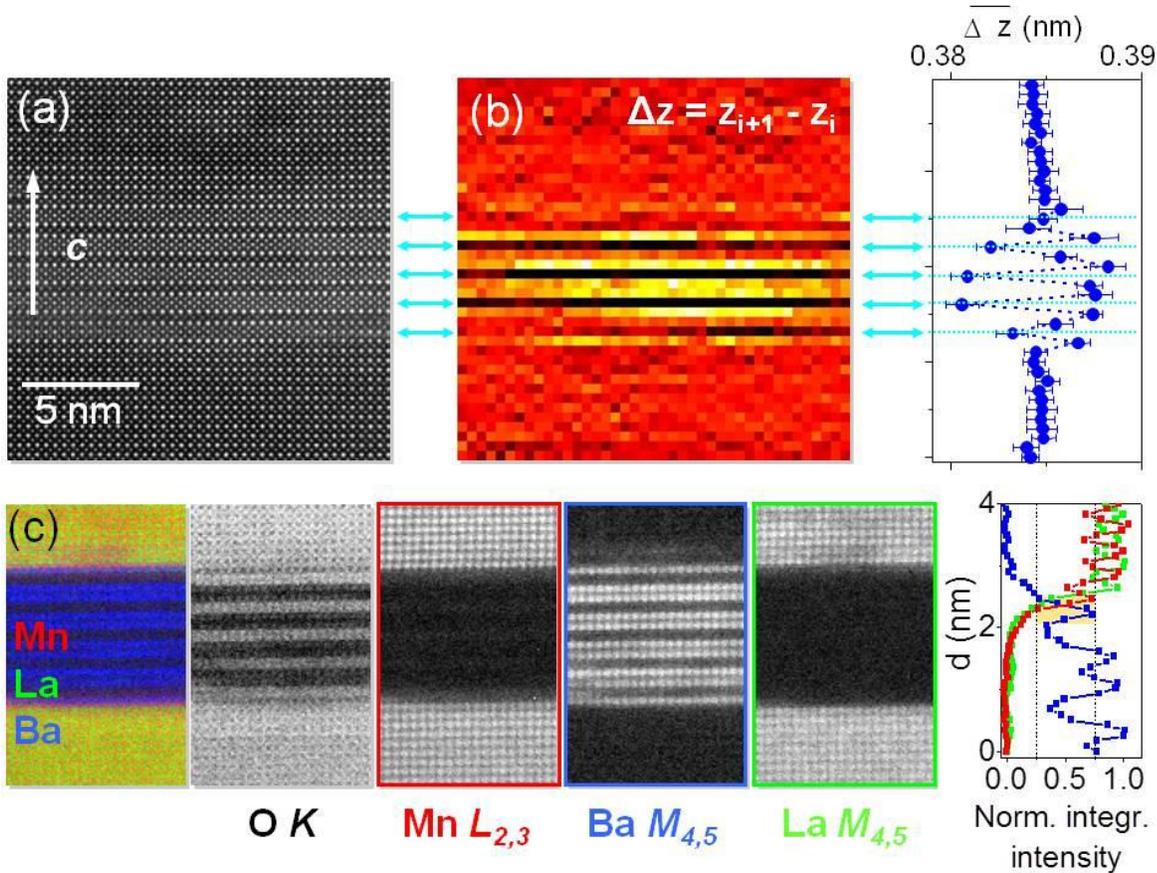

**FIG. 1:** (a) High resolution, Z-contrast image of a LCMO/YBCO/LCMO trilayer. (b) Map of transition metal spacings, $\Delta z$, along the c direction, with a lateral average of the image (right). $CuO_2$ biplanes are characterized by a smaller distance (dark stripes). (c) RGB compound image (left) and EELS maps of the integrated intensity of O K edge, Mn $L_{2,3}$, Ba $M_{4,5}$ and La $M_{4,5}$ edges, as labeled. The RGB imaged is obtained by overlaying the Mn (red), La (green) and Ba (blue) maps. The right panel shows the normalized integrated intensities of the Mn $L_{2,3}$ (red), Ba $M_{4,5}$ (blue) and La $M_{4,5}$ (green) across a LCMO(top)/YBCO(bottom) interface, extracted from an EELS linescan. An orange rectangle marks the width of a perovskite unit cell block at the interface.

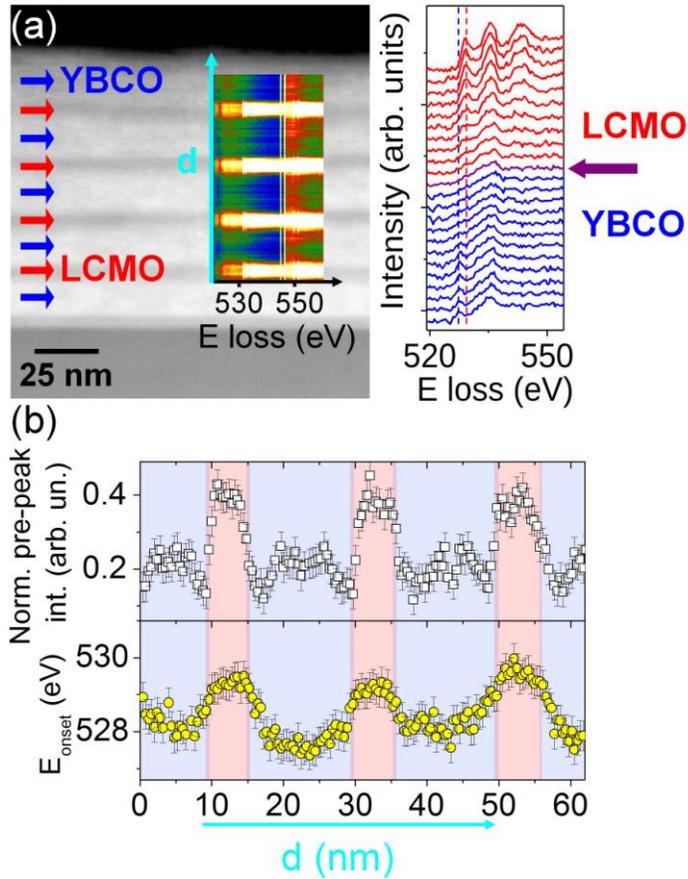

**FIG. 2:** (a) Z-contrast image of a LCMO/YBCO multilayer on a (100) $SrTiO_3$ substrate. Arrows mark LCMO (red) and YBCO layers (blue). The inset shows an EELS line-scan acquired along the growth direction. The right panel shows the energy range corresponding to the O K-edge across one of the YBCO-LCMO interfaces in the linescan. Dashed lines marked the position of the O prepeak for LCMO and YBCO away from the interface. (b) Prepeak intensity (top), and the edge onset position (bottom) along the growth direction, marked with a light blue arrow. Some data adapted from previous work (see supplemental materials for details).

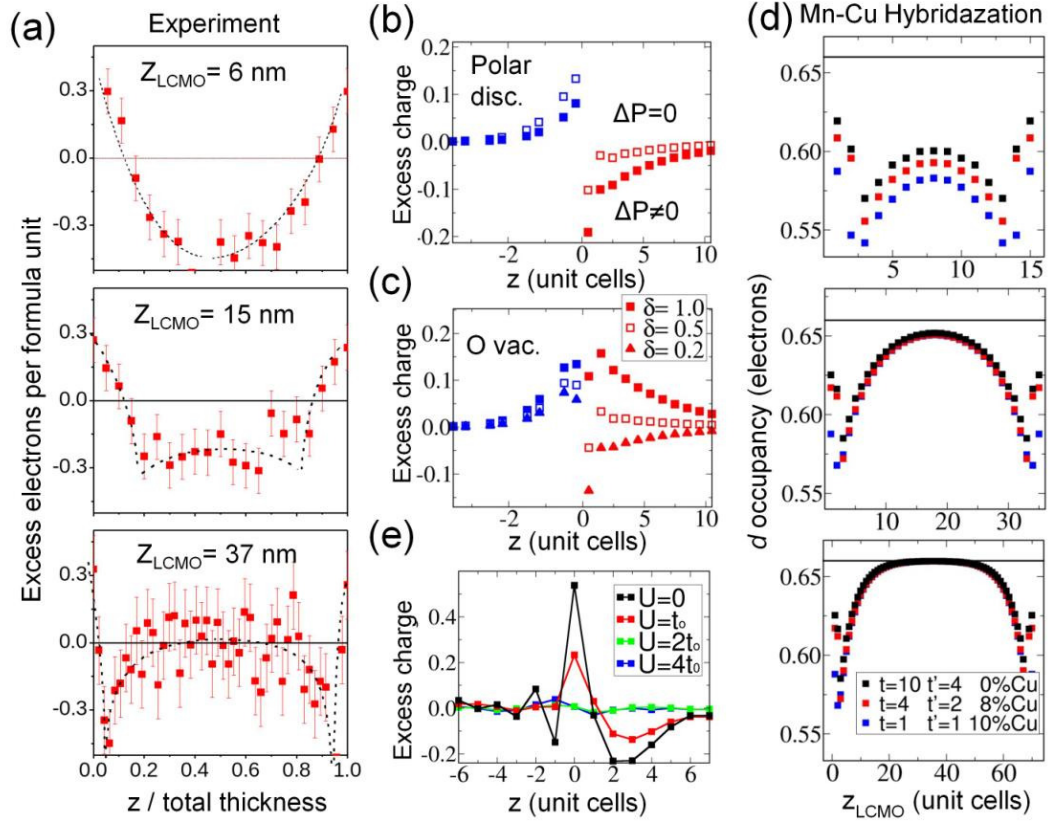

**FIG. 3**: Experimental and theoretical charge profiles. (a) Experimental charge profiles across the LCMO layer in multilayers with different thicknesses $Z_{LCMO}$. Some data adapted from previous work (see supplemental materials for details). (b) Results of the model calculations, including (empty symbols) and excluding (full) the polar discontinuity at the interface. Blue symbols correspond to YBCO and red symbols correspond to LCMO. Notice that polar discontinuity cannot account for the electron enrichment at LCMO near the interface. (c) Effect of oxygen vacancies. $\delta$ labels the oxygen deficiency in the Mn plane closest to the interface (of chemical formula $MnO_{2-\delta}$). Although oxygen vacancies dope the interface with electrons, the charge profiles in the LCMO layer decrease monotonically to zero, unlike the experiments in (a). (d) Effect of strong hybridization of Cu and Mn orbitals in the model and of hybridization together with Cu substitution in LCMO first atomic plane (as indicated), showing the non-monotonic charge profile as in (a). (e) Effect of Hubbard U interaction, showing a charge-transfer scenario for $U \leq t_0$. Details about the model in the main text and SM.

# Competition between covalent bonding and charge transfer at complex-oxide interfaces

Juan Salafranca,[1,2,*] Julián Rincón,[2,3] Javier Tornos,[1] Carlos León,[1] Jacobo Santamaria,[1] Elbio Dagotto,[2,4] Stephen J. Pennycook,[5,2] and Maria Varela.[2,1]

[1]Grupo de Física de Materiales Complejos, Universidad Complutense, 28040 Madrid, Spain

[2]Materials Science and Technology Division, Oak Ridge National Laboratory, Oak Ridge, TN 37831, USA

[3]Center for Nanophase Materials Sciences, Oak Ridge National Laboratory, Oak Ridge, TN 37831, USA

[4]Department of Physics and Astronomy, The University of Tennessee, Knoxville, TN 37996, USA

[5]Department of Physics and Astronomy, Vanderbilt University, Nashville, TN 37235, USA

*Correspondence to: jsalafra@ucm.es

# Supplemental Material

**Growth, electron microscopy and electron energy-loss spectroscopy**

High quality $La_{0.7}Ca_{0.3}MnO_3/YBa_2Cu_3O_{7-x}$ superlattices and trilayers were grown by high oxygen pressure sputtering as detailed in Ref. [1].

Data in Fig. 1 have been acquired in an aberration corrected Nion UltraSTEM100 operated at 100 kV equipped with a Gatan Enfina EEL spectrometer. The EELS maps were produced by integrating the intensity under the respective edges of interest after background subtraction using a power-law fit. Specimens were prepared by conventional methods (grinding and Ar ion milling). Principal component analysis was used to remove random noise. Data in Fig. 2 were acquired in an aberration corrected VG Microscopes HB501 UX equipped with a Gatan Enfina EELS, operated at 100 kV. The specimen was prepared by conventional methods. The prepeak normalized intensities were calculated by fitting Gaussian fits to the prepeak and the main peak and normalizing their respective areas (see Refs. [2-4]). Elemental traces in Fig. 1 have been produced by integrating a 30 eV wide window under the edge of interest, after removing the background using a power law fit. All signals decrease from 75% to 25% of their maximum signal within a unit cell, showing an atomically sharp interface. This technique, however, may not be adequate to quantify minor amounts of interface chemical disorder (in the range of a few percent units) due to the unavoidable presence of surface amorphous layers induced by the ion mill process.

Part of the data in Fig. 2(a), (b) and Fig. 3(a) is adapted from a previous, non-refereed publication [5] and also from Ref. [2] and [6].

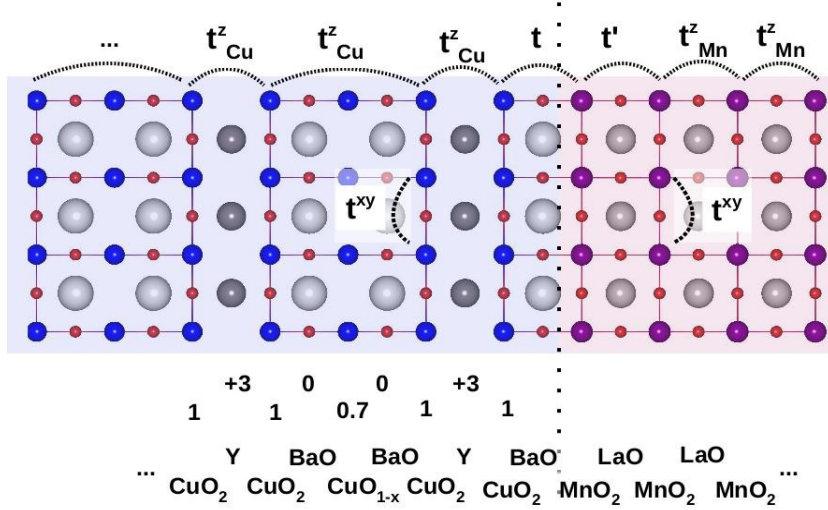

Fig. S1. Schematic illustration of the interface and the parameters used in the model. The *t's* indicate the different hopping parameters, including t and *t'* that are affected by increasing hybridization due to the presence of the interface. The numbers underneath each atomic plane represent the core charge (i.e. nuclei plus valence electrons).

**Model calculation details**

As stated in the main text, our goal is to calculate the charge distribution along the direction perpendicular to the LCMO/YBCO interface. Our model Hamiltonian for the interface includes effects such as interface polarity, oxygen vacancies, electron interactions, charge transfer relaxation, and chemical bonding. Only the $e_g$ orbitals are considered since the accepted view is that $t_{2g}$ orbitals are not active in LCMO or in YBCO. The Hamiltonian and the relevant parameters are explained next, and the Coulomb potential, $\phi_{COUL}$, is discussed below. A schematic illustration of the interface model and its parameters is shown in Fig. S1. Periodic boundary conditions are imposed in the (x, y) plane and open boundaries were considered for the z direction (perpendicular to the interface).

The LCMO/YBCO heterostructure Hamiltonian reads:

$$H = \sum e^i_\gamma n_{i,\gamma} + \sum t^{i,\alpha}_{\gamma\gamma'} C^\dagger_{i,\gamma} C_{i+\alpha,\gamma'} + h.c. + \phi_{COUL} \qquad (1)$$

The first two terms express the kinetic energy gain due to the hybridization of the d-symmetry orbitals in the usual nearest-neighbor tight binding form. $e_\gamma$ is the onsite energy, and $\alpha$ runs over the spatial directions, where z is the direction perpendicular to the interface. We assume x−y symmetry and translational symmetry of the in-plane unit cell. The electronically active sites are indexed by i. Following the widely accepted picture, we consider that the relevant orbitals in YBCO are in the $CuO_2$ planes, and that the $CuO_x$ chains act only as charge reservoirs. $\gamma$ and $\gamma'$ label the different orbitals.

As explained in the main text, the two $e_g$ orbitals are considered for both materials. For simplicity, we assume that $x^2 − y^2$ and $3z^2 − r^2$ are degenerate in LCMO, which is a good approximation at room temperature. In YBCO, due to the planar symmetry of Cu, $x^2 - y^2$ is lower in energy. The hopping parameters between the different orbitals along each of the three spatial directions are labeled by $t^\alpha$. Within each material, the relationship among the hopping parameters corresponding to different orbitals is dictated by symmetry [7,8]. In order to keep the number of parameters low, we choose $t^y$ $t^x$ to be the same in the cuprate and manganite side. For the z direction, $t_0 \equiv t_{3z^2-r^2, 3z^2-r^2}$ is the only non-zero element. The value of this parameter on the manganite side is the energy unit (approx. 0.5 eV [9]). $t_{3z^2-r^2, 3z^2-r^2}$ in the cuprate side is taken to be ten times smaller: the exact value is irrelevant for the physics discussed here, as long as it is significantly smaller than that of the manganite. The effective values of the hopping in the z direction at the interface, and in the first manganite layer, t and t', might be strongly affected by interface effects, such as lattice relaxations. Furthermore, in practice the effective $3z^2$-$3z^2$ hopping takes place via the apical oxygen, and a lower d-level energy would reduce the d-p "charge transfer" energy, increasing the effective hopping. As the proposed model is restricted to the d-orbitals, this effect can be described by the phenomenological parameters t, and t', at the interface. Therefore, we explore the effects of variations in these parameters in the main text.

In bulk YBCO, the $3z^2 - r^2$ orbital is completely full, while there are some holes in the $x^2 - y^2$ band ($e_{Cu}^{x^2-y^2} > e_{Cu}^{3z-r^2}$). The chemical potential of LCMO lies in the $e_g$ band, where both orbitals have very similar energies. Therefore, the difference ($e_{Mn}^{3z^2-r^2}$) $e_{Mn}^{x^2-y^2} - e_{Cu}^{x^2-y^2}$ is determined by the difference in (bulk) chemical potentials between the two materials. It is possible to estimate this difference from photoemission experiments ($\approx 0.85$ eV according to Ref. [10]). From our measurements of the core levels, it seems a factor of 2 larger. The surface dipoles introduce some uncertainty (measured work functions might depend on the particular surface termination [11]), and care should be paid when comparing these numbers with the shifts in absorption edges. We choose $e_{Mn}^{x^2-y^2} - e_{Cu}^{x^2-y\ 2} = 4t_0$ and $e_{Mn}^{x^2-y^2} - e_{Cu}^{3z^2-r^2} = 5t_0/2$ which produces a difference in bulk chemical potentials of the right order of magnitude and the proper ordering of the orbitals in the YBCO.

The long-ranged Coulomb potential at site i, between the ions and core electrons and the charge in the $e_g$ orbitals, in the Hartree approximation, takes the usual form:

$$\phi_{COUL} = \phi_i = \frac{1}{\epsilon_i} \sum_j \frac{\rho_i - \rho_j}{|r_i - r_j|} - \frac{1}{\epsilon_i} \sum_{\alpha, j \neq i} \frac{\rho_i - Z_\alpha}{|r_i - r_\alpha|} \quad . \tag{2}$$

$Z_\alpha$ is the charge of the ions plus the core electrons, i.e., all but the active $e_g$ electrons (see Fig. S1); $\rho_i$ is the remaining charge in the $e_g$ orbitals, determined self-consistently, $r_\alpha$ and $r_i$ are the positions of the nuclei and the center of the $e_g$ orbitals respectively, in units of the distance between atomic planes. $\epsilon_{Cu,Mn}$ are the effective dielectric constants for the two materials. They describe the electrostatic response of only the core electrons, and therefore their values are difficult to extract from experiments. We take $\varepsilon_{LCMO}=0.8$; $\varepsilon_{YBCO}=0.05$ (in units of $t_0^{-1}$) in order to fit the tails of the charge profile away for the interface in both materials. We notice that the presence of long range interactions is essential to the charge transfer phenomenon or semiconductor-like behavior.

The charge profile is obtained through a self-consistent procedure. Given a charge profile, the Coulomb potential is determined by resorting to Eq. (2), the new potential is inserted in Eq. (1), this equation is diagonalized in order to obtain a new charge profile, and the procedure is repeated self-consistently until convergence is reached. The calculations plotted in Fig. 3(d) were performed in trilayer consisting of 125, 165, and 235 atomic planes respectively. The larger trilayer consist of eight unit cells of YBCO, sixty two and half unit cells of LCMO and eight unit cells of YBCO. The number of layers of YBCO is the same in all trilayers, and enough to preclude surface-interface interactions. Temperature is imposed by filing the one-electron states according to a Fermi distribution. We choose $T=0.06t_0$ which roughly corresponds to room temperature.

The effect of polar discontinuity can be studied by properly altering the value of $r_\alpha$ in Eq. (2), placing the core and electronic charge in the corresponding positions. In this manner, only the net charge of each structural unit cell is taken into account, but not its formal dipole. The effect of oxygen vacancies is studied by changing systematically the value of the corresponding $Z_\alpha$ for the plane next to the structure, and monitoring the corresponding changes in the charge profile. The details of the charge distribution induced by oxygen vacancies – for the same vacancy concentration – are affected by the particular values of the hopping parameters. However, as stated in the main text, vacancies alone cannot produce a non-monotonic charge distribution without the competition of strong bonding and charge transfer across the interface.

The treatment of electron-electron interactions within Density Matrix Renormalization Group (DMRG) confines the calculation to one dimension (1D). This restriction does not alter the results as can be seen in Fig. 3 of the main text. Specifically for the one-orbital case, the electronic interaction considered here is the so-called Hubbard term, which is a local intra-orbital density-density interaction between electrons with opposite spin projections, namely,

$$H_U = \sum_i U_i \rho_{i,\uparrow} \rho_{i,\downarrow} \ . \qquad (3)$$

$U_i$ is the (in principle, material dependent) parameter controlling the electronic repulsion and $\rho_i$ is the charge density of the $e_g$ orbitals. For the two-orbital case, we have also included the effect of the Hund's ferromagnetic coupling, $J_H$, between spins in different orbitals, the inter-orbital Coulomb repulsion, and the so-called pairing hopping (see Ref. [8]). We take $J_H/U = 0.1$, for both manganite and cuprates, the specific values of U are indicated in Fig. 3(e) of the main text.

The effect of interaction is treated using DMRG, which has shown to be one of the most powerful methods to deal with strongly interacting systems in reduced dimensionality. The Hamiltonian implemented in DMRG is the 1D version of Eq. (1), i.e., only hopping along the z direction is taken into account; chemical potential unbalance, chemical bonding, and long range Coulomb repulsion are considered as well as in the original Hamiltonian model. As explained above, a self-consistent method was used in order to determine the charge profile; this procedure translates back to performing up to 600 sweeps in the finite-size algorithm in order to reach energy convergence. DMRG calculations were performed for both one- and two-orbital models, keeping up to 400 states per block with a truncation error around $10^{-7}$ for system sizes between 12 and 42 sites.

**Chemical disorder at the interface**

We have also studied the interdiffusion of the different cations across the interface in order to understand how they would affect the charge profile. La, Ca, Ba and Y are described in the model by their core charge. Y disorder has been completely ruled out by X-ray diffraction [12]. Disorder in the other elements, even up to our upper limit 10% substitution within the first unit cell, leads to very small changes in the charge profile, smaller than the data point sizes in Fig. 3.

Cu-Mn disorder is harder to include in the model. The core charges are the same for both, but the active orbitals near the Fermi energy have mainly Cu and Mn character. Therefore, the effect of Cu-Mn disorder in the band position and width should be larger than for the other cations. Since the core charge for Cu and Mn is the same, and the Cu-O-Cu and Mn-O-Mn hopping amplitudes are assumed to be similar for LCMO and YBCO, we focus on the effect of doping on the positions of the bands. In order to approximately quantify this effect we interpolate linearly the position of both $e_g$ orbitals ($e_{Mn}^{x2-y2}$, $e_{Cu}^{x2-y2}$) as a function of Cu content:

$$e_{(1-a)Mn + aCu}^{x2-y2} = (1-a)\, e_{Mn}^{x2-y2} + a\, e_{Cu}^{x2-y2}$$

An equivalent expression applies for the $3z^2 - r^2$ orbital. In this way, for 100% Cu substitution in a $MnO_2$ plane we would have the $e_g$ levels positioned as in YBCO. This is certainly a quite rough approximation only justified by the similarity in the electronic structure of different High Tc cuprates independently of the B site cations or the small differences in their atomic structure. Nevertheless, this linear approximation let us establish how chemical substitution affects the charge profile in a semi-quantitative way. The focus of the calculations is on the charge profile in the LCMO layers; therefore we restrict the following analysis to Cu substitution in the Mn position. It is expected that the largest substitution appears in the first layer of LCMO. For a particular amount of substitution x per Mn site in the first layer, we assume an $x^2$ substitution in the second layer, although substitution in the second layer is small and it barely affects the bands position (and therefore the resulting charge distribution) for any substitution values considered here.

Figure S2 illustrates the effect of 5% Cu substitution within the Mn site in the first plane of LCMO, which should already be detectable by the experimental techniques. The calculations in Fig. S2 show the theoretical charge profiles in LCMO when only substitution or hybridization (t=10, $t'$= 4) are taken into account, together with the profile for both mechanisms combined. It is important to remark that Cu substitution without the strong hybridization of Cu and Mn across the interface is not sufficient. In fact, no reasonable amount of Cu substitution alone is able to reproduce the features of the experimental profile.

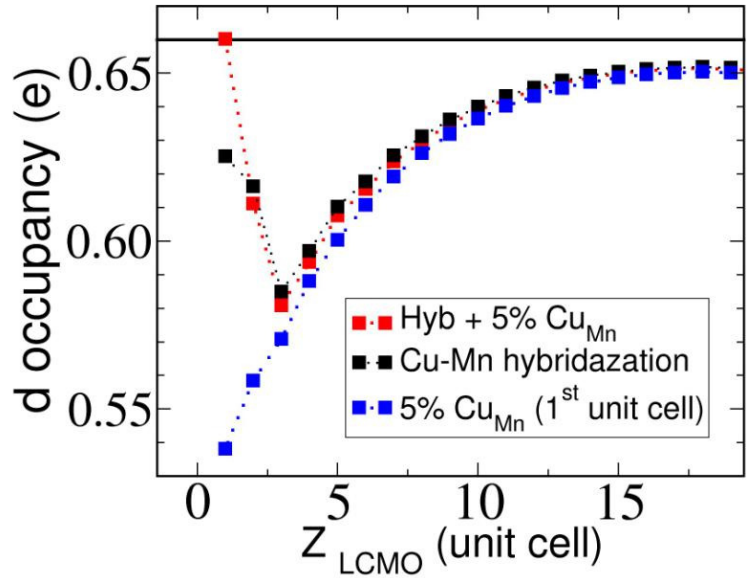

Fig. S2. Effect of different mechanisms considered in the model. Strong Cu-Mn hybridization (black) reproduces the non-monotonic behavior observed experimentally. When the effect of 5% Cu substitution into the Mn site of the first unit cell of LCMO is considered. The results (red) are even more similar to the experimental profile in Fig 3(a) of the manuscript. The same Cu substitution without the Cu-Mn hybridization (blue), however, is not able to reproduce the non-monotonic behavior. These results are partially reproduced in the mid panel of Fig. 3(d).